\documentstyle[twocolumn,aps,prl]{revtex}
\begin{document}


\title{Tensor Analyzing Powers for Quasi-Elastic Electron Scattering from
Deuterium}

\author{
Z.-L.~Zhou,$^{1,}$\cite{byline1}
M.~Bouwhuis,$^2$
M.~Ferro-Luzzi,$^{2,3}$
E.~Passchier,$^2$
R.~Alarcon,$^4$
M.~Anghinolfi,$^5$
H.~Arenh\"ovel,$^6$
R.~van~Bommel,$^2$
T.~Botto,$^2$
J.~F.~J.~van~den~Brand,$^{1,7}$
H.~J.~Bulten,$^{1,7}$
S.~Choi,$^4$
J.~Comfort,$^4$
S.~M.~Dolfini,$^4$
R.~Ent,$^8$
C.~Gaulard,$^4$
D.~W.~Higinbotham,$^9$
C.~W.~de~Jager,$^{2,}$\cite{byline2}
E.~Konstantinov,$^{10}$
J.~Lang,$^3$
W.~Leidemann,$^{11}$
D.~J.~de~Lange,$^2$
M.~A.~Miller,$^{1,}$\cite{byline3}
D.~Nikolenko,$^{10}$
N.~Papadakis,$^2$
I.~Passchier,$^2$
H.~R.~Poolman,$^2$
S.~G.~Popov,$^{10,}$\cite{byline4}
I.~Rachek,$^{10}$
M.~Ripani,$^5$
E.~Six,$^4$
J.~J.~M.~Steijger,$^2$
M.~Taiuti,$^5$
O.~Unal,$^1$
N.~Vodinas,$^2$
H.~de~Vries.$^2$\hfill}
 
\address{
$^1$\ Department of Physics, University of Wisconsin, Madison, WI 53706,
USA\\
$^2$\ NIKHEF, P.O. Box 41882, 1009 DB Amsterdam, The Netherlands\\
$^3$\ Institut f\"ur Teilchenphysik, Eidg. Technische Hochschule, CH-8093
Z\"urich, Switzerland\\
$^4$\ Department of Physics, Arizona State University, Tempe, AZ 85287, USA\\
$^5$\ Istituto Nazionale di Fisica Nucleare (INFN), I-16146 Genova, Italy\\
$^6$\ Institut f\"ur Kernphysik, Johannes Gutenberg Universit\"at, D-6500 Mainz,
Germany\\
$^7$\ Department of Physics and Astronomy, Vrije Universiteit,
1081 HV Amsterdam, The Netherlands\\
$^8$\ TJNAF, Newport News, VA 23606, and Department of Physics, Hampton
University, Hampton, VA 23668, USA\\
$^9$\ Department of Physics, University of Virginia,
Charlottesville, VA 22901, USA\\
$^{10}$\ Budker Institute for Nuclear Physics, Novosibirsk, 630090 Russian
Federation\\
$^{11}$\ Dipartimento di Fisica and INFN (gruppo collegato di Trento), Universit\`a di Trento, I-38050 Povo (Trento),
Italy\\
}
\date{\today}
\maketitle

\begin{abstract}
We report on a first measurement of tensor analyzing powers 
in quasi-elastic
electron-deuteron scattering at an average three-momentum transfer of 
1.7 fm$^{-1}$.
Data sensitive to the spin-dependent nucleon density in the deuteron
were obtained for missing momenta up to 150 MeV/$c$
with a tensor polarized $^2$H target internal to an electron storage ring.
The data are well described by a calculation that includes the effects of
final-state interaction, meson-exchange and isobar currents, and leading-order
relativistic contributions.
\end{abstract}
\bigskip

The deuteron is often used as a benchmark to test nuclear theory, 
since reliable calculations can be performed in both 
non-relativistic and relativistic models. 
Observables such as its binding energy, static magnetic dipole and
charge quadrupole moment, asymptotic $D/S$ ratio, and the elastic
electromagnetic form factors
place strong constraints on models for the nuclear interaction.
Furthermore, the deuteron is one of the few nuclear systems for which
predictions based on quantum chromodynamics have been made and
tested \cite{Brod73,Belz95,Carl90,Boc98}.
The understanding of the deuteron has progressed to a level which
allows the deuteron to be used as an effective neutron target in
studies of the electromagnetic form factors of 
the neutron \cite{Plat90,Eden94,Kle97,Gm}
and the spin structure functions of the neutron in deep inelastic
scattering \cite{SMC,E143,HERMES}.


Quasi-elastic scattering has been applied to determine
the momentum distribution of the proton in the deuteron \cite{Bern81}
with necessary corrections for the interaction effects \cite{Aren82}, 
and to study the $(e,e^\prime p)$ reaction mechanism itself \cite{Gilad}:
the necessity of including relativistic corrections has been demonstrated
by determining the longitudinal-transverse interference structure function;
details of the electromagnetic coupling to the nuclear currents
have been studied by performing longitudinal-transverse separations 
and polarization transfer measurements;
and the influence of final-state interaction (FSI) effects 
has been isolated by studying the fifth structure function.
%
%
However, all of these measurements were performed with unpolarized deuterons.
The use of polarized deuterium is advantageous, since one
can select polarization observables that are sensitive to small amplitudes
that are related to, for example, the charge form factor of the
neutron, or the $D$-wave admixture in the ground-state wave function
\cite{Aren95}.  Here, we report on the first measurement of quasi-elastic
electron scattering from polarized deuterium. 

It is well known that the tensor force has 
an important influence on deuteron structure. 
It introduces $D$-state components into 
the predominantly $S$-wave ground-state wave function 
and hence leads to a density distribution that depends
on the spin projection $m_z$.
Due to the tensor part of the nucleon-nucleon
interaction and the repulsive core at short distances,
the deuteron exhibits a toroidal shape for $m_z = 0$ and 
a dumbbell shape for $m_z = \pm 1$ (see e.g. \cite{Fore96}).
The spin-dependent momentum density distribution $\rho_{m_z}$ can be written as

\begin{eqnarray}
\label{rTd}
{\rho_0 ({\rm\bf p}) \over 4 \pi} &=& \{ R_0 + \sqrt{2} R_2 d^2_{0,0}(\theta)
\}^2 + 3 (R_2 d^2_{1,0}(\theta))^2 \hfill\nonumber\\
{\rho_{\pm 1}({\rm\bf p}) \over 4 \pi} &=& \{ R_0 - {1 \over \sqrt{2}} R_2
d^2_{0,0}(\theta) \}^2 + {9 \over 8} R_2^2 (1-\cos^4\theta), \hfill
\end{eqnarray}
\noindent

\noindent
with $R_0 (p)$  and $R_2(p)$ the usual radial wave functions 
for orbital angular momentum $L = 0$ and $L = 2$ in momentum 
space, respectively, $d^j_{m,m^\prime}$ the rotation functions
and $\theta$ the angle between the polarization axis and the
relative momentum ${\bf p}$ of the two nucleons.

The spin-dependent density distribution can be probed by 
quasi-elastic electron scattering from tensor-polarized deuterium targets,
which yields data complementary to 
the elastic channel \cite{The91,Fer96,CEBAF}
where the {\em total} nuclear current (i.e. nucleon and meson
contributions) is probed.
In plane-wave impulse approximation (PWIA), 
the $(e,e^\prime p)$ cross section for unpolarized electrons
factorizes in a part depending on 
the off-shell electron-proton cross section
and a part containing the spin-dependent momentum distribution as shown
in Eq. \ref{rTd}{\cite{CabD93}.
In that case the relative nucleon momentum ${\bf p}$ can be related
to the missing momentum of the reaction, defined as
${\bf p_m} = {\bf q} - {\bf p^\prime}$, with ${\bf q}$ the momentum transferred
in the electron scattering process and ${\bf p^\prime}$ the momentum
of the knocked-out proton.
In a cross section measurement with tensor-polarized deuterons one can define
the tensor analyzing power $A^T_d$:
\begin{eqnarray}
\label{aTd}
A^T_d&=&\sqrt{1 \over 2} {\sigma_+(p_m) + \sigma_-(p_m) -2\sigma_0(p_m)
\over \sigma_+(p_m) +
\sigma_-(p_m) +\sigma_0(p_m)} \hfill\nonumber\\
&{\stackrel{PWIA}{=}}& -{2 R_0(p) R_2(p) + \sqrt{1\over 2}
R_2^2(p) \over R_0^2(p) + R_2^2(p)} d^2_{0,0}(\theta),
\end{eqnarray}
\noindent
with $\sigma_0$ ($\sigma_\pm$) the cross section measured
with $m_z = 0$ ($m_z = \pm 1$) and 
$d^2_{0,0} = {3 \over 2} \cos^2\theta - {1 \over 2}$.
In the limit where $R_2\ll R_0$, $A^T_d$ directly 
relates to the ratio $R_2/R_0$. 
Furthermore, the extreme values of $A^T_d$ are 
$+\sqrt{1\over 2}$ (for $R_0 = -\sqrt{2} R_2$) and $-\sqrt{2}$
(for $R_0 = \sqrt{1 \over 2} R_2$) with $\theta = 0$. 
Under these conditions, in PWIA $\sigma_0 = 0$ ($\sigma_\pm = 0$)
and consequently, even when FSI and all other effects are included, 
the cross section is expected to be small for electron scattering 
from deuterons with the spin projection $m_z = 0$ ($m_z = \pm 1$).

The experiment was performed with a polarized gas target \cite{ABS} internal to
the AmPS electron storage ring at NIKHEF. A beam current of 120 mA was
injected. The 565 MeV electron beam had a lifetime of about 15 minutes.
An atomic beam source was used to inject a flux of
$1.1 \times 10^{16}$ atoms/s with two hyperfine states 
into a T-shaped storage cell, which was cooled to
100 K in order to further increase the target density. 
The data were obtained with Teflon-coated aluminum cells 
(wall thickness 25 $\mu$m) with diameters of 15 mm or 20 mm, 
and a length of 400 mm.
The resulting target thickness amounted up to 
$2 \times 10^{13}$ $^2$H$\cdot$atoms/cm$^2$.
The tensor polarization of the target $P_{\rm zz}$ ($= 1 - 3 n_0$, with $n_0$
the fraction of deuterons with $m_z = 0$) was varied every 10 seconds between
$P_{\rm zz}^+ = +0.488 \pm 0.014 \pm 0.03$ and $P_{\rm zz}^- =
-0.893 \pm 0.027 \pm 0.052$, where the first (second) error represents the
statistical (systematic)
uncertainty. The tensor polarization of the deuterium atoms was measured
{\sl in situ} by a polarimeter \cite{polar} that analyzes the fraction of
the target gas that has been ionized by the traversing electron beam.
Two electromagnets were used to generate a magnetic field in the 
interaction region that allows one to orient the target polarization axis in the
electron scattering plane, either parallel or perpendicular to the
momentum transfer.  A set of scrapers was employed to intercept
the halo of the electron beam and thus reduce the amount of
background from the cell wall.

The experimental setup has been described in detail in 
Refs \cite{ABS,detect},
and results for elastic electron-deuteron scattering have been presented
previously \cite{Fer96}; here only a brief overview is given.
Electrons scattered from the tensor-polarized deuterium gas
were detected in an electromagnetic calorimeter, consisting of 6 layers
of CsI(Tl) blocks covering a solid angle of 180 msr. Two plastic scintillators,
one in front of the CsI(Tl) blocks, one positioned between the first two
layers, provided the electron trigger. The total energy resolution obtained
(about 22 MeV) was sufficient to distinguish between events from 
quasi-elastic scattering and events from pion electroproduction.
Two sets of wire chambers, one adjacent
to the scattering chamber, one in front of the first trigger scintillator,
were used for track reconstruction. The central angle of the electron detector
was 35$^\circ$ corresponding to an average transferred
three-momentum of $\vert {\bf q} \vert \approx 1.7$ fm$^{-1}$.

The ejected protons were detected in a range telescope, consisting of 15
layers of 1 cm thick plastic scintillator preceded by a layer of 2 mm thick
plastic scintillator. The trigger was formed by a coincidence between the first
two layers. The detector was positioned
at a central angle of 80$^\circ$, and covered a solid angle of nearly 300 msr.
The range telescope was preceded by two wire chambers for track
reconstruction. Protons in the range of 30-100 MeV were detected
with an energy resolution of about 1.5 MeV.

Calibration measurements were performed with the kinematically overdetermined
reaction $^1$H$(e,e^\prime p)$. Since the calorimeter
excludes the pion-production channels, the five-fold differential
cross section can be obtained by measuring the electron scattering
angles $\theta_e , \phi_e$, the angles of the ejected proton
$\theta_p , \phi_p$ and the energy of the proton. This procedure
resulted in an excellent missing-momentum resolution, 
for a non-magnetic detector setup, of 6.2 MeV/$c$ as
demonstrated in Fig. \ref{eepspec} (top panel). 
In the bottom panel of the figure, the
missing-momentum spectrum for scattering from deuterium is shown.
The data are in
reasonable agreement with the result of a Monte Carlo calculation
that includes the off-shell electron-proton cross section of de
Forest Jr. \cite{Forest}, momentum densities from Bernheim $et$ $al.$
\cite{Bern81} and a model for the detector phase space. The shaded
histogram represents the background contribution which was determined
by scattering from an empty cell. 
It is seen that the background contributes significantly 
to the data taken for missing momenta above 150 MeV/c.

Fig. \ref{asymspinang} shows the measured asymmetry, $A^T_d$,
 as a function of the
angle $\theta_s$ between the polarization axis and the missing momentum.
Results of measurements with the polarization axis parallel and
perpendicular to the momentum transfer were combined in this figure.
The data are compared to predictions \cite{Aren95} for the Paris potential.
The dashed curve in Fig. \ref{asymspinang} represents
the results for the plane-wave Born approximation (PWBA)
which includes the coupling to the neutron, whereas in the solid
curve the effects of final-state interaction, meson-exchange (MEC) and
$\Delta$-isobar (IC) currents, and relativistic corrections are included.
The spread in kinematics in each bin has been taken into account by
applying a Monte-Carlo code that interpolated between a dense grid of
calculations that covered the full acceptance of our setup.
It is observed that the calculations describe the data well.
Since in PWIA the asymmetry is proportional to $d^2_{0,0}$ (see Eq. \ref{aTd}),
the zero crossings of the asymmetry are predicted to be at
$\cos \theta_s = \pm \sqrt{1 \over 3}$ ($\approx 0.58$),
while without a $D$-state component to the
ground-state wave function the asymmetry vanishes.
The results for PWBA differ slightly from this prediction, indicating that
for our kinematics the coupling to the neutron
is  not completely negligible.
Furthermore, it is demonstrated that FSI 
effects modify the asymmetry mainly for positive values of $\cos \theta_s$.

Fig. \ref{asympm} shows $A^T_d$ as a function of missing momentum in
approximately parallel kinematics, i.e. the center-of-mass angle 
between ${\bf p^\prime}$ and
${\bf q}$ has been restricted ($\theta_{pq}^{cm} < 13^\circ$). 
The results of the full calculation describe the data well. 
The inclusion of spin-dependent rescattering effects improves the description
at the lowest missing momenta ($p_m \simeq 50$ MeV/$c$), where $D$-state
contributions are small.
Note that, the tensor analyzing power is sizable, even at the relatively
small missing momenta addressed in the present experiment.
This is due to the fact that
an interference between the small $D$-state amplitude and the dominant $S$-state
amplitude is probed (see Eq. \ref{aTd}), whereas in the unpolarized cross
section only the sum of the squares of the $S$-state
and $D$-state amplitudes enter.

We have investigated the influence of the choice of nucleon-nucleon
potential on the calculated tensor analyzing powers. The results for $A^T_d$ 
calculated with the Paris, Bonn, Nijmegen, and 
Argonne potentials differ by less than 1\% for missing momenta
below 150 MeV/$c$ and therefore do not influence our conclusions. In addition,
we studied the influence of relativistic effects in the current operator.
$A^T_d$ changes by less than 2\% when the model is modified
to not include third- and fourth-order terms in $p/M_p$ in the current
operator and to neglect contributions from the boost operator that translates
the initial-state wave function from the laboratory frame to the
center-of-mass frame.

In summary, a polarized deuterium target has successfully been utilized
in quasi-elastic electron scattering. Tensor analyzing powers have been
measured for missing momenta up to 150 MeV/c. 
The data are well described by the present model \cite{Aren95}, 
using the Paris potential, and the tensor analyzing
power is dominated by the effects of the $D$-state contribution.
We have shown that the internal-target technique allows for 
novel measurements of quasi-free scattering from the deuteron, 
which in the future should be extended to higher missing momenta,
in the region where the tensor analyzing powers are predicted
to strongly depend on the details of 
the deuteron spin structure \cite{Fore96,Fran83,Bian96}.

\medskip
We gratefully acknowledge discussions with S. Jeschonnek and T. W. Donnelly
on factorization issues of the spin-dependent (e,e$^\prime$p) cross section.
This work was supported in part by the Stichting voor Fundamenteel Onderzoek
der Materie (FOM), which is financially supported by the Nederlandse
Organisatie voor Wetenschappelijk Onderzoek (NWO), the Swiss
National Foundation, the National Science
Foundation under Grants No. PHY-9316221 (Wisconsin), PHY-9200435 (Arizona
State) and HRD-9154080 (Hampton),
NWO Grant No. 713-119, and HCM Grants No. ERBCHBICT-930606 and ERB4001GT931472.

\begin{figure}
\caption{Top panel: missing-momentum spectrum for the kinematically
overcomplete reaction $^1$H$(e,e^\prime p)$. 
Bottom panel: missing-momentum spectrum for scattering from
deuterium. The shaded histogram indicates the background from the cell wall.
The solid histogram shows a Monte Carlo prediction. \hfill}
\label{eepspec}
\end{figure}

\begin{figure}
\caption{ $A^T_d$ as a function of $\cos \theta_s$. The dashed curve
represents the result for PWBA, the solid curve for the full calculation
including FSI, MEC, and IC contributions and relativistic corrections.}
\label{asymspinang}
\end{figure}

\begin{figure}
\caption{$A^T_d$ as a function of $p_m$ for parallel kinematics
(i.e. $\theta_{pq}^{cm} < 13^\circ$). The short-dashed curve represents the
result for PWBA, in the long-dashed curve also FSI-effects are
included, and the solid curve represents the full calculation.
\hfill}
\label{asympm}
\end{figure}
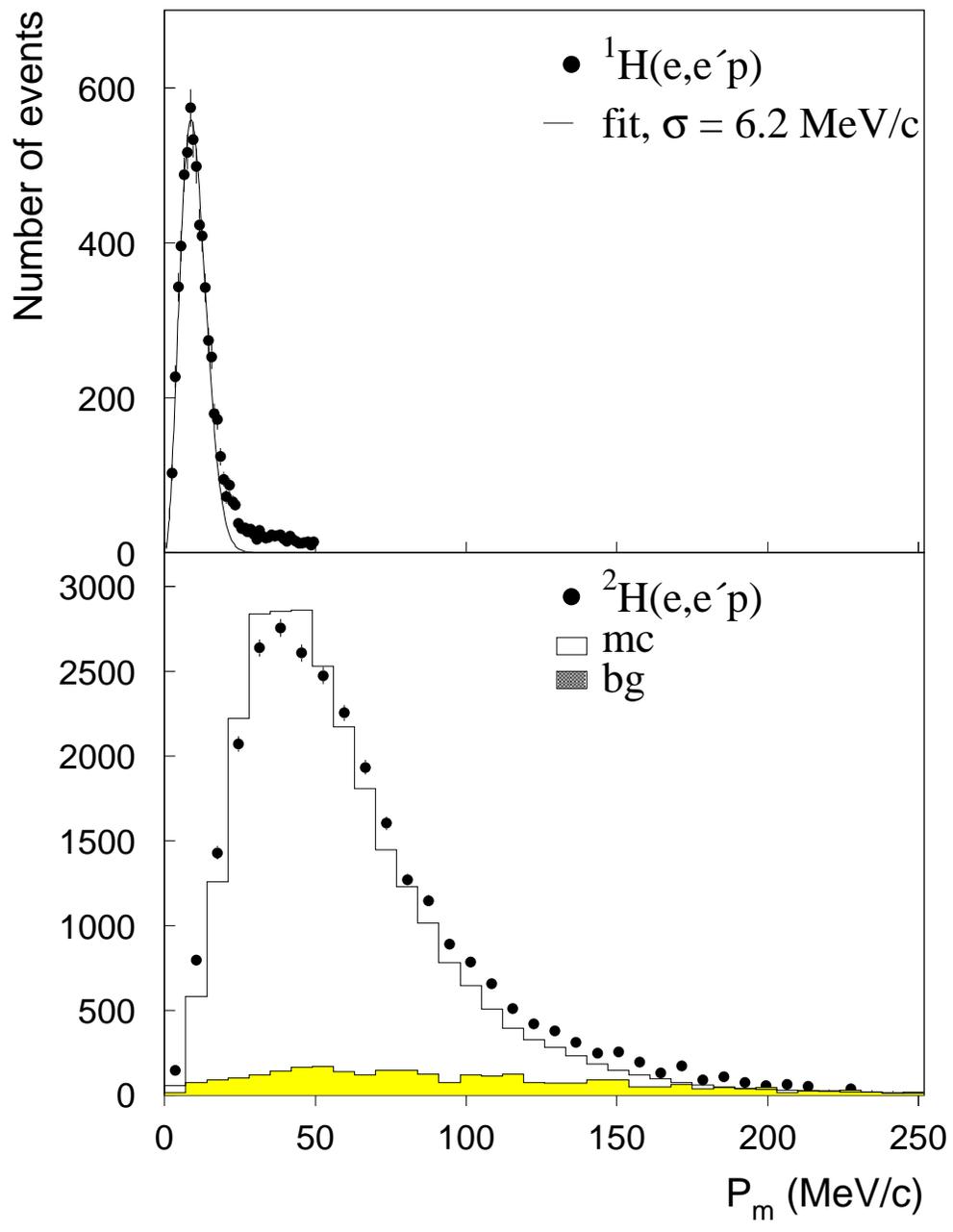
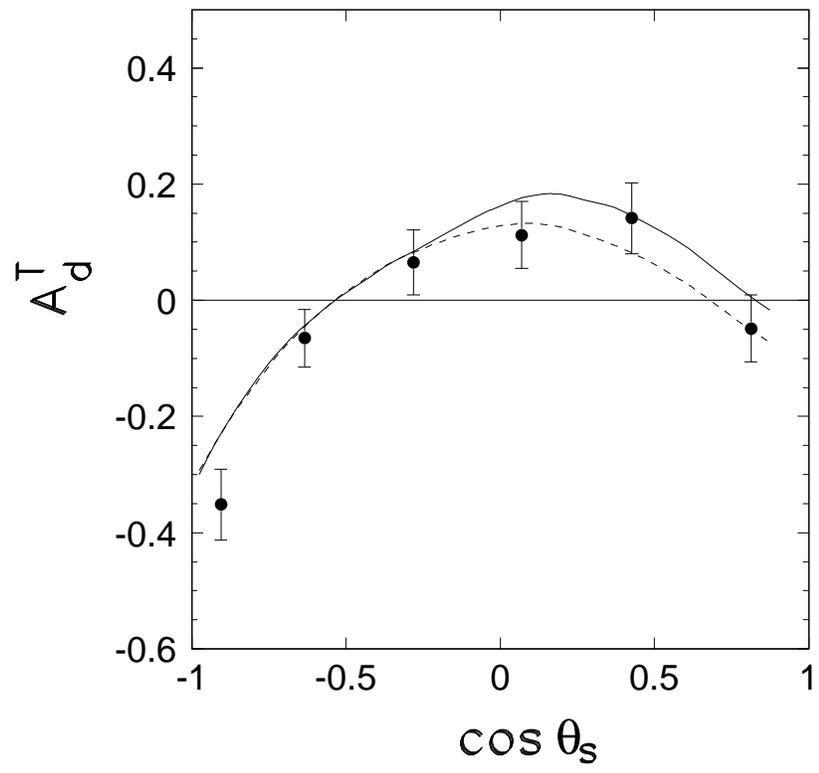
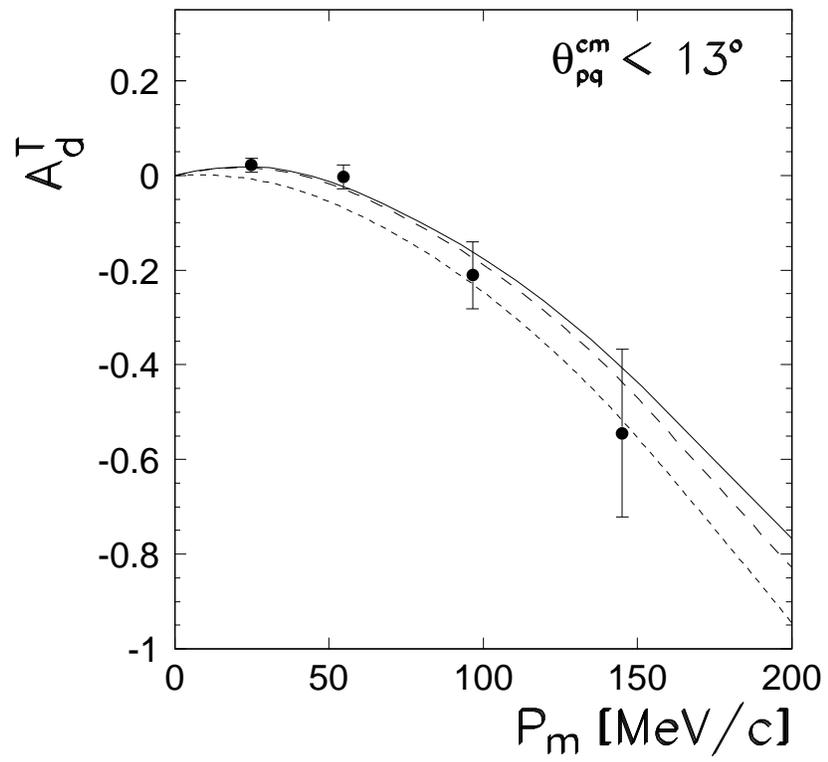
\end{document}